\documentclass[aps,prl,twocolumn,showpacs,superscriptaddress,groupedaddress]{revtex4}

\usepackage[utf8x]{inputenc}
\usepackage{amsmath,amssymb,amsfonts,subfigure,braket,color,dsfont}
\usepackage[english]{babel}
\usepackage{graphicx} %(if I use above command, I can't resize figures....)
\usepackage{sistyle}
\usepackage{listings}
\usepackage{gensymb} %to use \degree

\begin{document} 
 
\title{Inhomogeneous response of an ion ensemble from mechanical stress}
\author{S. Zhang} 
\affiliation{LNE-SYRTE, Observatoire de Paris, Universit\' e PSL, CNRS, Sorbonne Universit\' e, Paris, France}
\author{N. Galland} 
\affiliation{LNE-SYRTE, Observatoire de Paris, Universit\' e PSL, CNRS, Sorbonne Universit\' e, Paris, France}
\affiliation{Univ. Grenoble Alpes, CNRS, Grenoble INP and Institut N\' eel, 38000 Grenoble, France}
\author{N. Lu{\v c}i\'c} 
\affiliation{LNE-SYRTE, Observatoire de Paris, Universit\' e PSL, CNRS, Sorbonne Universit\' e, Paris, France}
\author{R. Le Targat} 
\affiliation{LNE-SYRTE, Observatoire de Paris, Universit\' e PSL, CNRS, Sorbonne Universit\' e, Paris, France}
\author{A. Ferrier} 
\affiliation{Chimie ParisTech, Universit\' e PSL, CNRS, Institut de Recherche de Chimie Paris, 75005 Paris, France} 
\affiliation{Sorbonne Universit\'e, Facult\'e des Sciences et Ing\'enierie, UFR 933, 75005 Paris, France} 
\author{P. Goldner} 
\affiliation{Chimie ParisTech, Universit\' e PSL, CNRS, Institut de Recherche de Chimie Paris, 75005 Paris, France} 
\author{B. Fang} 
\affiliation{LNE-SYRTE, Observatoire de Paris, Universit\' e PSL, CNRS, Sorbonne Universit\' e, Paris, France}
\author{Y. Le Coq}
\affiliation{LNE-SYRTE, Observatoire de Paris, Universit\' e PSL, CNRS, Sorbonne Universit\' e, Paris, France}
\author{S. Seidelin}\email{signe.seidelin@neel.cnrs.fr}
\affiliation{Univ. Grenoble Alpes, CNRS, Grenoble INP and Institut N\' eel, 38000 Grenoble, France}
\affiliation{Institut Universitaire de France, 103 Boulevard Saint-Michel, F-75005 Paris, France}

\date{\today}

\begin{abstract}

Material strain has recently received growing attention as a complementary resource to control the energy levels of quantum emitters embedded inside a solid-state environment. Some rare-earth ion dopants provide an optical transition which simultaneously has a narrow linewidth and is highly sensitive to strain. In such systems, the technique of spectral hole burning, in which a transparent window is burnt within the large inhomogeneous profile, allows to benefit from the narrow features, which are also sensitive to strain, while working with large ensembles of ions. However, working with ensembles may give rise to inhomogeneous responses among different ions. We investigate experimentally how the shape of a narrow spectral hole is modified due to external mechanical strain, in particular, the hole broadening as a function of the geometry of the crystal sites and the crystalline axis along which the stress is applied. Studying these effects are essential in order to optimize the existing applications of rare-earth doped crystals in fields which already profit from the more well-established coherence properties of these dopants such as frequency metrology and quantum information processing, or even suggest novel applications of these materials, for example as robust devices for force-sensing or highly sensitive accelerometers.
 
\end{abstract}

%modify PACS
\pacs{42.50.Wk,42.50.Ct.,76.30.Kg}

\maketitle

The extraordinary coherence properties of rare-earth ion dopants in a host crystal have led to their description in terms of a ``frozen ion gaz'' as they constitute the solid state emitters that come the closest to isolated ions trapped with electromagnetic fields inside a vacuum chamber in terms of coherence properties~\cite{Yano1991,Equall1994,Thiel2011,Goldner2011}. However, the impact of the crystalline matrix cannot be entirely neglected, as illustrated for instance by the large inhomogeneous absorption profile of the ion ensemble, arising from a non-uniform distribution of static strain due to the doping which slightly distorts the crystalline lattice locally. The technique of spectral hole burning has been shown to surmount such inhomogeneous effects, as ions with a particular frequency are selectively pumped to an hyperfine, dark state. The width of the created spectral hole, despite the fact that a large number of ions participate in the signal, reflects the single-ion characteristics, and as long as the crystal is not subject to external perturbations, these characteristics are preserved in time. This has been extensively exploited for instance in ultra-high-precision laser stabilisation and spectroscopy~\cite{Julsgaard2007,Thorpe2011,Gobron2017}, in which a laser is frequency stabilized by locking it to a narrow spectral hole. However, if the crystal containing a spectral hole is subject to modifications in its physical environment (for instance, external stress), the ions constituting the spectral hole may react differently according to their local environment, and in addition to a global displacement of the spectral hole, a deformation can occur. In order to fully benefit from solid state systems based on ensembles, such effects must be studied and understood.

In this work, we are particularly interested in the 580 nm optical transition in $\rm Eu^{3+}$ ions in an $\rm Y_2SiO_5$ (YSO) host matrix, as it is both ultra-narrow linewidth, potentially down to 122 Hz~\cite{Equall1994}, and is sensitive to mechanical strain, allowing for an efficient strain coupling between dopants and crystalline matrix. Earlier experiments have demonstrated such strain coupling in the context of optomechanical systems, for instance using quantum dots~\cite{Yeo2014} or NV centers in diamond~\cite{Teissier2014,Ovartchaiyapong2014,Macquarrie2017}, but these emitters either exhibited a large linewidth or low strain sensitivity. Generally speaking, optical transitions exhibit relatively high strain sensitivity, but in most solid state emitters, such transitions exhibit very broad (in the GHz range) linewidths, with, at best, zero-phonon lines in the MHz range. On the contrary, the electronic structure of trivalent lanthanide ions, in which the optical transitions inside the 4f shell are protected by 5s and 5p fully populated shells, allows for record narrow transitions, as well as the high strain sensitivity that is typical for an optical transition. Thus, when viewing the strain sensitivity as a resource, this kind of system represents the best of both worlds, as witnessed by proposals for creating strain-engineered quantum hybrid systems based on this system~\cite{Molmer2016,Seidelin2019}. On the other hand, depending on the application, this high sensitivity can also be a nuisance, of which the effect needs to be minimized. For instance, vibrations of the sample give rise to an acceleration of the crystal which leads to a compression, in turn creating fluctuations in strain \cite{Chauvet2019}, which can limit the performance of the system. Such vibrations often occur due to the operation of a cryostat, and represents a recurring challenge for many applications with this type of dopants. Therefore, whether a resource or a nuisance, an in-depth study of the response of $\rm Eu^{3+}$ in a YSO matrix to strain seems valuable in a variety of contexts.

\begin{figure}[t]
\centering
\includegraphics[width=80mm]{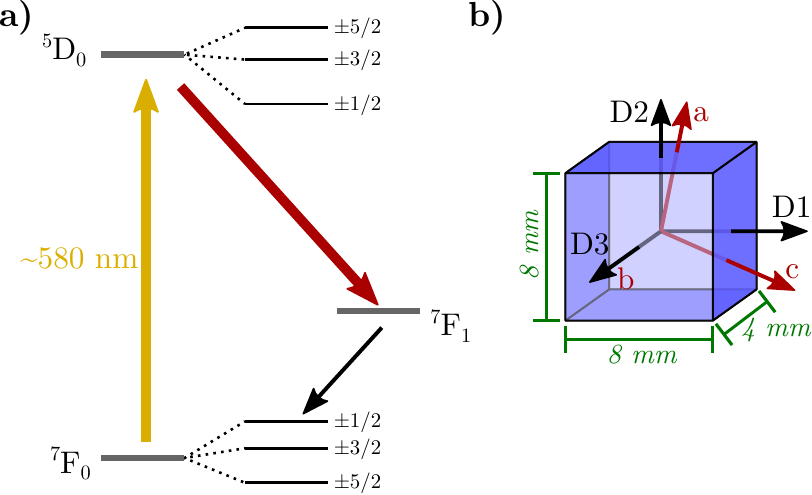}
\caption{\label{levelsCrystal} Details on ion level structure and crystal axes. In a) the energy diagram of the $\rm Eu^{3+}$ ions, where each electronic level is separated in 3 hyperfine states. When excited at resonance at 580 nm, the ions decay radiatively primarily to $^7F_1$ and then non-radiatively to one of the hyperfine states. In b) we show the  $\rm Y_2SiO_5$ crystal, which is cut parallel to the 3 dielectric axes, D$_1$, D$_2$ and D$_3$, as well as its dimensions.}
\end{figure}

We study the $^7F_0 \rightarrow$  $^5D_0$ transition shown in figure~\ref{levelsCrystal} a). More precisely, the YSO crystal possesses two locations within the unit cell where $\rm Eu^{3+}$ can substitute for $\rm Y^{3+}$, referred to as site 1 and 2, with vacuum wavelengths of 580.04\,nm and 580.21\,nm, respectively. These are the only two non-equivalent substitution sites in the crystalline matrix, both exhibiting a $C_1$ symmetry. Each state is composed by 3 hyperfine states, with mutual separations in the 30-100 MHz range~\cite{Yano1991}. As illustrated in figure~\ref{levelsCrystal} a), the spectral holes are formed by pumping the ions resonantly from the $^7F_0$ to the $^5D_0$ state. From there, they decay radiatively, mainly to the $^7F_1$ state \cite{Konz2003}, before decaying non-radiatively to the three hyperfine states in the $^7F_0$ manifold. Optical pumping prevents population build-up in the hyperfine level with which the pump beam is resonant, creating a transparent window in the inhomogeneous profile at this exact frequency. A spectral hole represents approximately $10^{13}$ $\rm Eu^{3+}$ ions.
 
We recently determined the frequency shift of this transition by studying the linear displacement in frequency of a spectral hole when applying external stress to the crystal~\cite{Galland2019}. Here, we will instead focus on how the shape of the spectral hole evolves according to the applied stress, for these two different crystal sites, as this provides insight into the microscopic environment of the ion dopants constituting the spectral hole. As the direction parallel to which the stress is applied also plays an important role, in the overall frequency shift as well as in the broadening, we investigate the application of stress along two non-equivalent axes:  either parallel to the D$_1$ or to the D$_2$ dielectric axes, see figure~\ref{levelsCrystal} b). For maximum absorption, the pump beam is propagated along the crystalline $b$-axis (which coincides with the dielectric D$_3$ axis) and polarized linearly along the D$_1$ axis~\cite{Konz2003,Ferrier2016}. 

\begin{figure}[t]
\centering
\includegraphics[width=80mm]{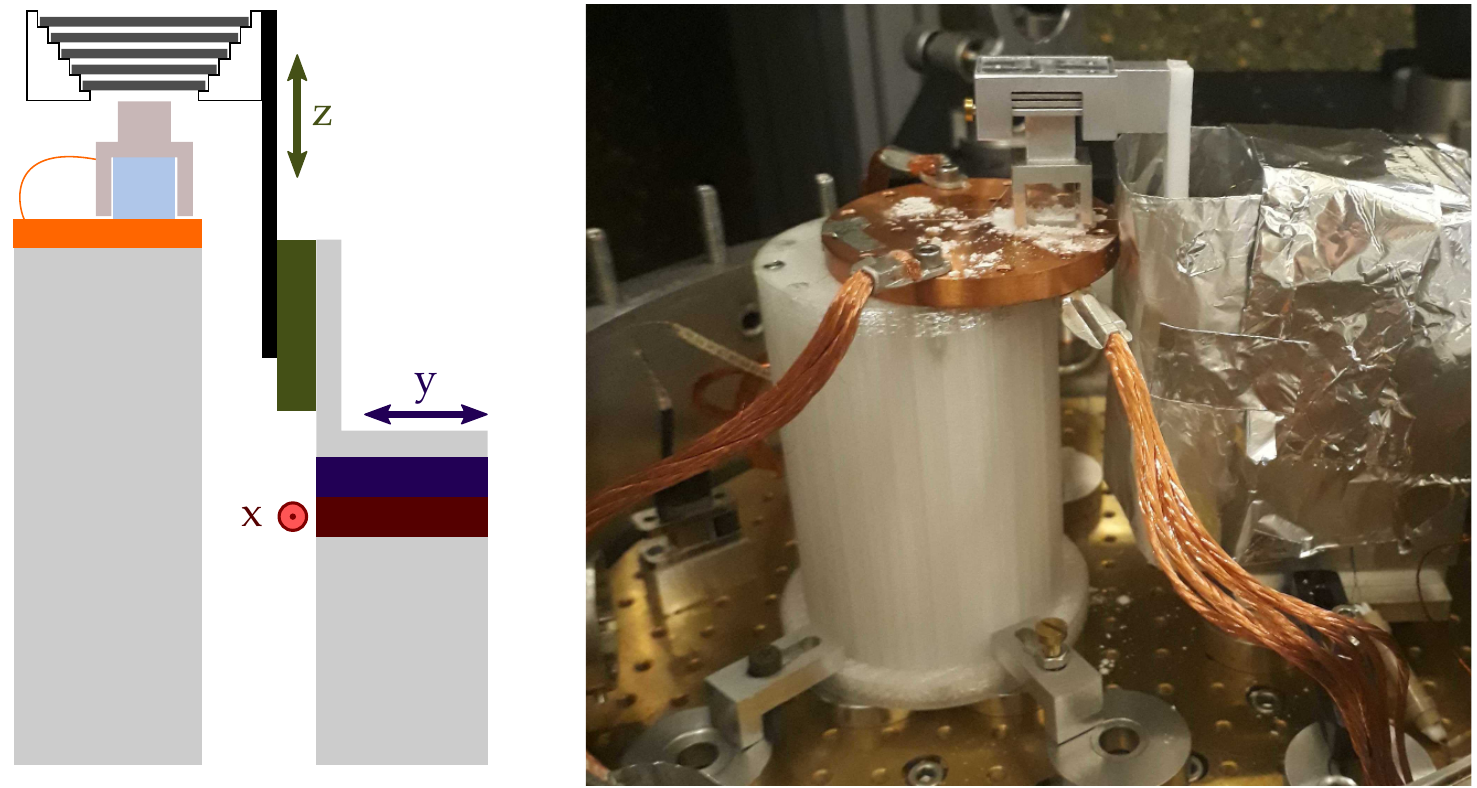}
\caption{\label{setup}Experimental device for applying a calibrated stress to the crystal. Left: A schematics of the ladder structure which allows us to deposit a given number of weights on the top of the crystal (illustrated by the light blue square), by vertically displacing the structure ($z$-axis translation stage), and to center the weights (by using the $x,y$-axis). Right: A photo of the device, mounted inside the cryostat science chamber. In both schematics and photo, an electrically grounded metal structure surrounding the crystal is visible, which prevents a build-up of static charges.}
\end{figure}

The physical setup is similar to the one described in detail elsewhere~\cite{Gobron2017,Galland2019}. In brief, we use 2 diode lasers (master and slave) at 1160 nm. The master laser is locked to a frequency stabilization cavity by the Pound Drever Hall method allowing for a frequency instability below $10^{-14}$ for time constants of 1-100\,s (corresponding to a few Hz linewidth). The slave laser is locked to the master laser.  They are both subsequently frequency doubled to reach 580 nm, with an output intensity of approximately 5 mW. The slave laser possesses the same stability as the master, but its frequency is continuously tunable in a range of several GHz.  In addition, acousto-optic modulators allow us to scan the frequency across the spectral structures with a range of approximately 1 MHz.  An absolute frequency measurement is provided by referencing the signal to a frequency comb. To monitor the shape of the spectral hole, we use an avalanche photo-diode to record the laser intensity after absorption of the crystal, normalized by the laser intensity send directly to the detector. Polarizing beamsplitters combined with $\lambda$/2 waveplates allows for adjusting the intensity. The crystal is maintained at 3.15 K in a closed-loop cryostat positioned on an active vibration isolation platform.

In order to quantitatively study the response of a spectral structure to mechanical stress, a calibrated force in a given direction must be applied at cryogenic temperature. We use a staircase-like structure (see figure~\ref{setup}), with a vertical position being controlled by a cryo-compatible motorized platform, providing a way of placing a variable number of weights on the top of the crystal inside the science chamber in the cryostat (see photo in figure~\ref{setup}). From the mass of the weights, we have access to an absolute calibrated force. We start the sequence by burning a spectral hole with no weights on the crystal, and we then add the number of weights corresponding to the desired stress. 
 
\begin{figure}[t]
\centering
\includegraphics[scale=1]{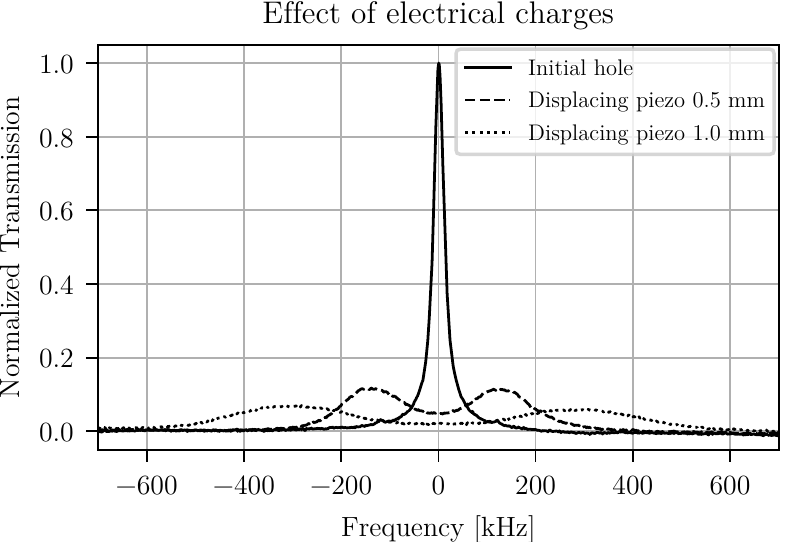}
\caption{\label{piezo} The effect on a spectral hole of operating the piezo translation stage near the crystal, for crystal site 1 (without applying weights) prior to electrical shielding. The solid line corresponds to the initial spectral hole, and the dashed and dotted line to the hole after having operated the piezo one and two steps, respectively. The deformations of the spectral holes persist in time, despite ended operation of the piezo translation stage.}
\end{figure}

During the initial stages of the experiment, we observed irreversible effects on the spectral hole while operating the translation stage for depositing the weights. More precisely, by operating the piezo translation stage, even without depositing weights on the crystal, the hole would exhibit a splitting, and moreover, the spectral hole would not return to its initial shape, even after the piezo operation has ended (note however that no permanent micro-structural modification of the crystal was observed). Examples of this behavior is illustrated in figure~\ref{piezo}, where we show an example of the shape of a spectral hole after having operated the piezo stage for a given excursion distance. This behavior can be understood as the energy transitions in $\rm Eu^{3+}$ are also sensitive to electric fields~\cite{Thorpe2013,Mcfarlane2014,Li2016}, and when operating the motorized translation stage (which works by stick-slip mechanism of a piezo on a fine threaded screw), accumulated stray charges can generate irreversible displacements or deterioration of the spectral hole. Moreover, silver lacquer on the crystal used for optimal thermal contact seemed to exert an additional stress on the crystal while applying weights, modifying the spectral hole in a manner depending of the amount and position of the silver lacquer in contact with the crystal. The solution we adopted consists in removing the silver lacquer and replacing it with a metal shield positioned on the top of, and extending down around, the crystal, but without direct contact with the vertical sides of the crystal. In addition we surrounded the entire piezo translation stage with a metal cover (visible in the photo in figure~\ref{setup}.) Both the crystal- and piezo stage cover were subsequently grounded. The weights for generating stress, instead of being deposited directly on the crystal, were then put on the crystal metal shield. These modifications resulted in the disappearing of any effects on shape of the spectral hole due to operating the piezo translation stage.

\begin{figure}[t] 
\centering
\includegraphics[scale=1]{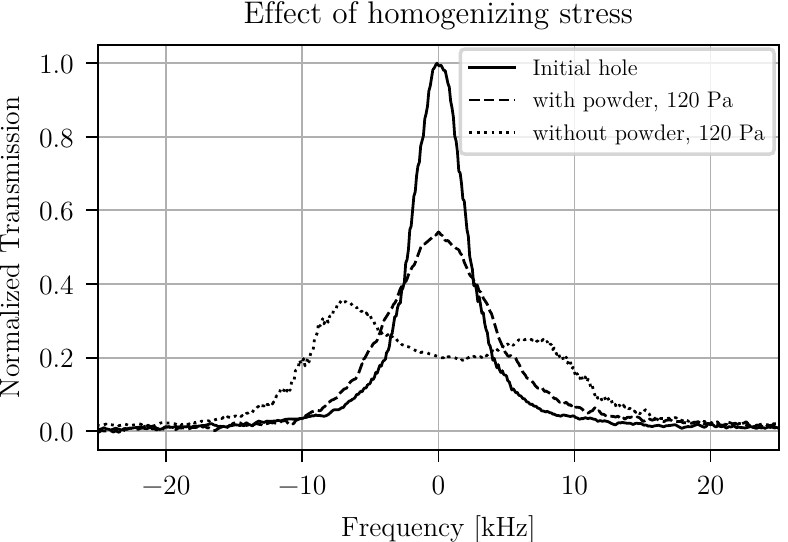}
\caption{\label{powder_spectrum} The effect of stress on the shape of the spectral hole, for crystal site 2, with stress applied along the D1 axis. The solid line shows the initial hole without stress. The dotted line corresponds to the case in which the stress is applied without depositing a crystalline powder at the contact surfaces, and the dashed line to the case in which a powder has been used. The stress also gives rise to a global shift in frequency of the spectral structures, but this shift has been subtracted in order to center the curves in frequency.}
\end{figure}

However, when subsequently adding weights to the crystal, instead of a pure shift in frequency of the spectral hole, we observed a clear splitting in the spectra. As such splitting is not expected to appear as long as the stress is applied perpendicular to the $b$-axis~\cite{Bartholomew2014} (which is our case, as we apply stress parallel to either D$_1$ or D$_2$), we suspected that the stress was not perfectly parallel to the axis chosen. One potential explanation was that the weights were not completely centered relative on the crystal. To improve on this aspect, an $x-y$ translation stage was added for centering the objects on the crystal, and the splitting seemed to decrease slightly, but was not eliminated. We suspected that the contact surface between the weights and crystal, despite all the care taken, was not sufficiently uniform as to create a strain field perfectly perpendicular to the $b$-axis inside the crystal. In order to homogenize the stress, we used a $\approx$ 1 mm layer of YSO powder with a grain size of the order of micrometers at the two contact surfaces of the crystal. That is, we deposited the powder on the top of the crystal  (between the crystal and the metal shield), and below the crystal (between the crystal and the cryostat platform). When repeating the experiment, we observed that the splitting is no longer visible for the given pressures. These effects are illustrated in figure \ref{powder_spectrum}, in which we have applied stress parallel to the D$_1$-axis. The solid line corresponds to the initial spectral hole with a width below 5 kHz. When applying stress, before homogenizing the stress with the crystalline powder, we obtain spectra exhibiting large splittings, as illustrated by the dotted line, which appear to be eliminated by the use of the crystal powder (dashed line).

%By using the crystal powder, the splitting disappears entirely, and we obtain structures as illustrated by the dashed line. 
 
\begin{figure}[t]
\centering
\includegraphics[width=80mm]{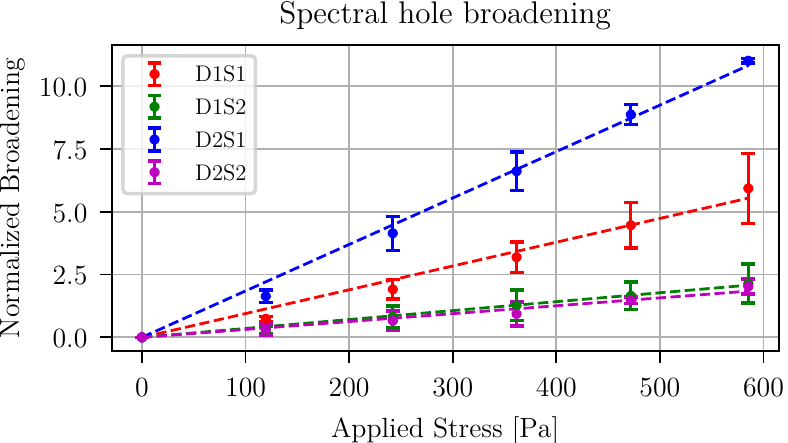}
\caption{\label{broadening} The effect of the stress on the width of a spectral hole, for different directions parallel to which the stress is applied (D1 and D2) and crystal sites (S1 and S2). The broadening has been normalized by the strain sensitivity, in order to plot $(\Gamma-\Gamma_0)/(\Gamma_0\lvert\kappa_{DiSj}\rvert/\kappa_{D1S1})$, where $\Gamma_0$ is the initial width of the unperturbed hole. The dashed lines are linear fits to the data.}
\end{figure}

We do observe, however, a persistent broadening of the spectral hole with applied stress, an effect we wish to investigate quantitatively in the following. First, in order to verify that the strain does not exhibit a significant variation across the crystal, after burning a spectral hole and adding strain, we probe the spectral hole in 5 different locations (center, left, right, top and bottom) of the crystal, with a beam diameter of 2 mm. The spectral hole appears to have the same shape in the different location, and the same global shift in frequency (with a variation much smaller than the width of the spectral hole), eliminating the possibility of a strong strain gradient across the crystal. In order to study the residual broadening systematically, we record the shape of the spectral hole as a function of stress applied (corresponding to 1 to 5 weights added to the crystal) with a 2.5 millimeter beam diameter, centered in the middle of the crystal. The experiment is repeated for both crystal sites (S1 and S2) and for adding the stresses parallel to the D$_1$ and D$_2$ axes. The result of this study is shown in figure \ref{broadening}. By denoting the initial width of the spectral hole, before applying stress, $\Gamma_0$, and the width measured for a given stress $\Gamma$, the broadening is given by $(\Gamma-\Gamma_0)/(\Gamma_0)$. For the data in figure \ref{broadening}, the initial hole widths are the following: $\Gamma_0$(D1S1)=5.4 kHz, $\Gamma_0$(D1S2)=4.0 kHz, $\Gamma_0$(D2S1)=3.8 kHz and $\Gamma_0$(D2S2)=4.4 kHz. We checked that the width of the initial hole does not change (to within the uncertainty of the determination of the width of the hole) in the temperature range between 3,1 K and 4,2 K. As our actual temperature fluctuations are at the 1 mK level, the width of the initial hole should therefore remain constant during the measurement sequence. This is further confirmed by removing all weights at the end of the measurement sequence, and observing that the hole regains its initial shape.

Moreover, in the case that the broadening arises from a residual inhomogeneity in the amplitude of the stress field in the crystal across the optical spot, the broadening should depend directly on the strain sensitivity, which is different for different crystal sites and axes parallel to which the stress is applied. That is, the same gradient in strain should give rise to a larger broadening for a more sensitive site and direction of stress. Thus, in order to compare the different measurements, we have also normalized the broadening according to the sensitivity. We previously determined these values~\cite{Galland2019}: $\kappa_{D1S1}=46\pm17$ Hz Pa$^{-1}$;  $\kappa_{D1S2}=137\pm16$ Hz Pa$^{-1}$; $\kappa_{D2S1}=-19\pm10$ Hz Pa$^{-1}$; and $\kappa_{D2S2}=-213\pm13$ Hz Pa$^{-1}$, where we have used the notation $\kappa_{DiSj}$ where $Di$ corresponds to the direction parallel to which the strain is applied, and $Sj$ the crystal site ($i,j$ =1 or 2). Using these values, we choose to normalize the broadening to the sensitivity of the crystal site 1, with a stress applied parallel to the D$_1$ axis. That is, we plot $(\Gamma-\Gamma_0)/(\Gamma_0\lvert\kappa_{DiSj}\rvert/\kappa_{D1S1})$. The results are shown in figure \ref{broadening}. Here, the error-bars account for the reproducibility of the experiment (4 to 6 measurements per data point), and solid lines correspond to linear fits to the data. By allowing for a quadratic component, we notice a small improvement of the quality of the fit, but the linear contribution is clearly dominant, so in what follows, we will consider this component only.
 
In the simple case where the broadening were due uniquely to an inhomogeneity in the amplitude of the crystal strain, we would expect all linear fits to exhibit the same slope to within the error-bars, as we have normalized the broadening to the strain sensitivity. However, we observe a significant variations in slopes: $\alpha_{D1S1}=
9.5\pm 4.6$ kPa$^{-1}$,  $\alpha_{D1S2}=3.6\pm2.1$ kPa$^{-1}$; $\alpha_{D2S1}=18.5\pm0.2$ kPa$^{-1}$, and $\alpha_{D2S2}=3.1\pm0.1$ kPa$^{-1}$. The uncertainties on the slopes account for the repeatability of the experiment. Even taking into account the uncertainty on both the $\kappa_{DiSj}$ and the $\alpha_{DiSj}$ coefficients, $\alpha_{D2S1}$ cannot be brought to overlap with $\alpha_{D1S2}$ and $\alpha_{D2S2}$. We cannot exclude a small inhomogeneity along the axis of propagation of the probe beam (along the $b$-axis). However, a gradient in the amplitude of strain in this direction would also give rise to a broadening being identical (when normalized by the sensitivity) among sites and stress-directions. In order to explain the variations of the slopes due to the inhomogeneity in the strain, one would require a component of the force to be in a direction perpendicular to the applied stress-direction. This is thus unlikely to account for the differences observed.

Instead, we may, at least partially, attribute the observed effect to an inhomogeneity of the different ions' responses on a microscopic scale, as suggested by other authors for a different system~\cite{Reeves1989}. In our case, the ions forming the spectral hole all have a slightly different electromagnetic environment, coming from a non-uniform distribution of static strain within the crystal which stems essentially from local distortion of crystal lattice due to doping, which also accounts for the large inhomogeneous profile. Although the ions contributing to the spectral hole all have the same absorption frequency (to within the width of the hole), this frequency is both a combination of the exact hyperfine states between which they have transitioned, which are separated by several tens of MHz (all three hyperfine levels are populated at 4 K prior to optical pumping), combined with the local static crystal strain. That is, contributing to a single spectral hole, there will be ions with internal static strain varying both in amplitude and direction, and they are therefore likely to react differently according to the externally applied strain. Although the strain sensitivities cited above have been shown to be independent on the frequency position in the inhomogeneous profile~\cite{Galland2019}, this explanation can still be valid: In each position of the profile, a spectral hole will be formed by ions with different local strain environment (reacting differently to strain), but the global strain sensitivity corresponds to an average effect of the ensemble of the ions, which can still be identical to that of a second position in the inhomogeneous profile. 

Finally, another type of inhomogeneity on a microscopic level could stem from the fact that transitions among different hyperfine states could react differently to strain as a direct consequence of atomic structure, but comparing with other systems, we believe the corresponding variation may be negligible. For instance, for the NV$^-$ in diamond, the strain-sensitivity of the optical 638 nm transition is 1000 Hz Pa$^{-1}$~\cite{Davies1976} while for the ground-state spin-levels, the sensitivity only amounts to 0.01 Hz Pa$^{-1}$~\cite{Ovartchaiyapong2014}, thus $10^5$ times smaller. Of course, dedicated measurements will be required in order to confirm whether this also applies to the case of rare-earth ion dopants, but the absence of distinct components in the profile of the spectral hole subject to stress also supports this assumption.

In conclusion, we have discussed how the shape of a narrow spectral hole is sensitive to stray charges, and we have shown how to eliminate these detrimental effects. We have also demonstrated that adding a fine layer of crystalline powder at the contact interfaces of a crystal significantly reduces inhomogeneities, both in direction and amplitude, in the strain field generated in the crystal due to added weights. This is a prerequisite for studying the inhomogeneities of the ion dopants on a microscopic level. The fact that the broadening, normalized by the strain-sensitivity, is distinct for different crystal sites and directions for the stress, indicate that the various ions participating in the spectral structure react differently according to their local strain environment. Future work includes repeating the experiments with a much smaller number of ions to see how this influences the inhomogeneous effects. Ideally, one would study the response to strain of a single ion, but currently, the smallest number of Eu$^{3+}$ ions in an YSO matrix detected is approximately 10~\cite{Casabone2018}, made possible by enhancing the signal by means of a micro-cavity, and thus currently incompatible with the application of an external, calibrated stress. However, by using a protocol of hyperfine class selection already demonstrated for rare-earth ion dopants \cite{Lauritzen2012} it should be possible to assess to which extend different classes react differently to strain. Theoretical calculations of the frequency shift of an ion dopant as a function of its local strain environment would also be of great value to better understand inhomogeneous effects resulting in spectral hole broadening, but due to the lack of symmetry of the crystal and the substitution sites, such calculations are extremely challenging. Even if the exact positions of the individual atoms can be calculated under strain, the calculation of Eu$^{3+}$ energy levels under strain would still have to be performed. This is currently out of reach of the best crystal field models. At present, it is therefore not possible to propose a model describing the results, thus the measurements seem to be the most efficient avenue for investigating these effects.

 %This inhomogeneous response can be due to a variation of the applied field across the crystal, but can also be due to an inhomogeneous response of

\vspace{0.2 cm}
 
The project has been supported by the European Union's Horizon 2020 research and innovation program under grant agreement No 712721 (NanOQTech). It has also received support from the Ville de Paris Emergence Program, the R\'{e}gion Ile de France DIM C'nano and SIRTEQ, the LABEX Cluster of Excellence FIRST-TF (ANR-10-LABX-48-01) within the Program ``Investissements d'Avenir'' operated by the French National Research Agency (ANR), and the EMPIR 15SIB03 OC18 and from the EMPIR program co-financed by the Participating States and from the European Union's Horizon 2020 research and innovation program.

\end{document}